\documentclass{article}
\usepackage{amssymb}
\usepackage{amsmath}
\usepackage{palatino}

\numberwithin{equation}{section}
\numberwithin{figure}{section}

\newcommand{\benumerate}{\begin{enumerate}}
\newcommand{\eenumerate}{\end{enumerate}}

\newcommand{\bitemize}{\begin{itemize}}

\newcommand{\eitemize}{\end{itemize}}
\newcommand{\der}[2]{\frac{\partial #1}{\partial #2}}

\newcommand{\dersec}[2]{\frac{\partial^{2} #1}{\partial #2^{2}}}

\newcommand{\dermixd}[3]{\frac{\partial^{2} #1}{\partial #2 ~\partial #3}}

\newcommand{\simbon}[2]{\mathop{\rm #1}\limits_{#2}}

\begin{document}


\title{Dispersionless integrable equations as coisotropic deformations.
Extensions and reductions}

\author{B.G. Konopelchenko$^{1}$ and F. Magri$^{2}$
\\
$^{1}$\small{Dipartimento di Fisica, Universita' di Lecce and
Sezione
INFN,}\\  \small{73100 Lecce, Italy}\\
$^{2}$\small{Dipartimento di Matematica ed Applicazioni,
Universita' di Milano Bicocca,}\\ \small{20126 Milano, Italy}}

\date{}
\maketitle

\begin{abstract}
Interpretation of dispersionless integrable hierarchies as
equations of coisotropic deformations for certain associative
algebras and other algebraic structures is discussed. It is shown
that within this approach the dispersionless Hirota equations for
dKP hierarchy are nothing but the associativity conditions in a
certain parametrization. Several generalizations are considered.
It is demonstrated that the dispersionless integrable hierarchies
of B type like the dBKP hierarchy and the dVN hierarchy represent
themselves the coisotropic deformations of the Jordan's triple
systems. Stationary reductions of the dispersionless integrable
equations are shown to be connected with the dynamical systems on
the plane completely integrable on a fixed energy level.
\end{abstract}


\newpage

\section{Introduction to the concept of coisotropic deformations.}
\label{sec_intro} Dispersionless integrable equations and
hierarchies have attracted considerable interest during the last
two decades (see e.g. [1-16]). A principal example of such
equations is given by the dispersionless Kadomtsev-Petviashvili
(dKP) equation
\begin{gather}
\label{dKP}
\begin{aligned}
\der{u}{x_{3}} &= \frac{3}{2} u \der{u}{x_{1}} + \der{v}{x_{2}}, \\
\der{v}{x_{1}} &= \frac{3}{4} \der{u}{x_{2}}.
\end{aligned}
\end{gather}
This equation is equivalent to the compatibility condition
\[
\dermixd{S}{x_{2}}{x_{3}} = \dermixd{S}{x_{3}}{x_{2}}
\]
for the pair of equations
\begin{align} \label{lax_dKP1} \der{S}{x_{2}} &= \left(\der{S}{x_{1}} \right)^{
2} + u, \\
\label{lax_dKP2} \der{S}{x_{3}} &= \left(\der{S}{x_{1}}
\right)^{3} + \frac{3}{2} u \der{S}{x_{1}} + v.
\end{align}
The higher dKP equations arise as the compatibility conditions for
equation~(\ref{lax_dKP1}) and equations
(see~\cite{Za1}-\cite{Za3})
\begin{equation}
\label{lax_dKPhigh} \der{S}{x_{n}} = \left(\der{S}{x_{1}}
\right)^{n} + \sum_{m=0}^{n-2} u_{nm}(x) \left(\der{S}{x_{1}}
\right)^{m}.
\end{equation}
Other dispersionless integrable equations are usually associated
with the Hamilton-Jacobi type equations of the
form~\cite{Kri3,Za3}
\begin{equation}
\label{lax_HJ} \der{S}{x_{n}} = \Omega_{n} \left(\der{S}{x_{1}} ,x
\right)
\end{equation}
where $\Omega_{n}$ are meromorphic functions of the first
argument.

In this standard formulation the variables $x_{1}$ and $p =
\der{S}{x_{1}}$ play a distinguished role. They form the pair of
canonically conjugate variables and the compatibility conditions
for the equations~(\ref{lax_HJ}) can be written as
\begin{equation}
\label{omega_comp} \der{\Omega_{n}}{x_{m}}\left(p,x \right)
-\der{\Omega_{m}}{x_{n}}\left(p,x \right) -
\left\{\Omega_{n}\left(p,x \right), \Omega_{m}\left(p,x \right)
\right\} = 0
\end{equation}
where the Poisson bracket $\left\{\cdot,\cdot \right\}$ is defined
as
\begin{equation}
\label{poisson_def} \left\{f,g \right\} = \der{f}{x_{1}}
\der{g}{p} - \der{f}{p} \der{g}{x_{1}}.
\end{equation}
So the whole standard approach deals with the nonstationary
Hamilton-Jacobi equations~(\ref{lax_HJ}) and the Hamiltonians
$\Omega_{n}(p,x)$.

In the present paper we will consider a novel approach to the
dispersionless integrable equations and hierarchies in which they
represent  themselves the coisotropic deformations of associative
algebras or other algebraic structures~\cite{KM}.

This approach which is a melting of ideas borrowed from
Hamiltonian mechanics and theory of associative algebras starts
with the reinterpretation of the Hamilton-Jacobi
equations~(\ref{lax_HJ}) as the stationary ones. Within such a
viewpoint it is natural to consider all variables
$x_{1},x_{2},x_{3},\dots$ on the equal footing and to introduce
the corresponding canonical variables $p_{1},p_{2},p_{3},\dots$
conjugate to $x_{1},x_{2},x_{3},\dots$ with respect to the
standard Poisson bracket $\left\{\cdot,\cdot,\right\}$ given by
\begin{equation}
\label{poisson_defgen} \left\{f,g \right\} = \sum_{i=1}^{n} \left(
\der{f}{x_{i}} \der{g}{p_{i}} - \der{f}{p_{i}}
\der{g}{x_{i}}\right).
\end{equation}
Thus we introduce the symplectic manifold $M^{2n}$ equipped with the
Poisson bracket~(\ref{poisson_defgen}). Hamilton-Jacobi
equations~(\ref{lax_HJ}) are substituted by the zero set of
Hamiltonians
\begin{equation}
\label{H_zeroset} h_{n} \equiv -p_{n} + \Omega_{n} \left(p_{1},x
\right) = 0.
\end{equation}
We emphasize that now $x_{i},p_{i}$ form pairs of canonical
conjugate variables and $p_{i} \neq \der{S}{x_{i}}$. We will
denote the submanifold $M^{2n}$ given by zero
set~(\ref{H_zeroset}) as $\Gamma$.

For instance, instead of
equations~(\ref{lax_dKP1}),~(\ref{lax_dKP2}) one now has the zero
set of the Hamiltonians
\begin{gather}
\label{dKP_zeroset}
\begin{aligned}
h_{2} &= -p_{2}+ p_{1}^{2}+ u \left(x_{1},x_{2},x_{3} \right) ,\\
h_{3} &= -p_{3}+ p_{1}^{3}+ \frac{3}{2} u p_{1} + v.
\end{aligned}
\end{gather}
The submanifold $\Gamma$ defined by the
equations~(\ref{dKP_zeroset}) is ``parametrized" by the functions
$u$ and $v$ of the coordinates $x_{1}$, $x_{2}$, $x_{3}$.

The second step in our approach is to introduce a special class of
submanifolds $\Gamma$. It is quite natural to require that
Hamiltonian flows generated by the Hamiltonians $h_{2}$ and $h_{3}$
preserve $\Gamma$ or, in other words, the Hamiltonian vector fields
associated with $h_{2}$ and $h_{3}$ are tangent to $\Gamma$. This
requirement is equivalent to the condition
\begin{equation}
\label{tang_h2h3} \left .\left\{h_{2},h_{3} \right\}
\right|_{\Gamma} = 0
\end{equation}
It is easy to check that this condition is satisfied if $u$ and $v$
obey the dKP equation~(\ref{dKP}). In this case $\left\{
h_{2},h_{3}\right\} = 0$.

Thus, for any solution $u$, $v$ of the dKP equation~(\ref{dKP})
the Hamiltonians $h_{2}$ and $h_{3}$ are in involution and the
deformation of $x_{1}$, $x_{2}$, $x_{3}$ according to the
Hamiltonian flows preserve $\Gamma$.

We will refer to such deformations as the coisotropic deformations
due to the following reason. The submanifold $\Gamma$ defined by
the equations~(\ref{dKP_zeroset}) and obeying the above conditions
is the $4-$dimensional submanifold in $M^{6}$ for which $\left.
\left\{h_{2},h_{3} \right\} \right|_{\Gamma} = 0$ and moreover the
restriction of the standard symplectic $2-$form
\begin{equation}
\label{twoform} \omega \equiv \sum_{i=1}^{3} dp_{i} \wedge dx_{i}
\end{equation}
on $\Gamma$ does not vanish. Since the Hamiltonian vector fields
associated with $h_{2}$ and $h_{3}$ span the kernel of $\omega$
then the rank of $\left.\omega \right|_{\Gamma}$ is equal to two
and, hence, \begin{equation} \label{twoform_G} \omega_{\Gamma} =
d{\cal L} \wedge d{\cal M}
\end{equation}
where ${\cal L}$ and ${\cal M}$ are canonical variables on
$\Gamma$. The submanifolds with such properties are called
coisotropic submanifolds (see e.g.~\cite{Wein1,Ber}).

So, any solution $u$, $v$ of the dKP equation defines
through~(\ref{dKP_zeroset}) a $4-$dimensional coisotropic
submanifold in $M^{6}$ and it is then natural to refer to the
associated deformations as coisotropic ones.

Similarly, the conditions $\left\{h_{n},h_{m} \right\} = 0$ for
the Hamiltonians $h_{n}$ given by~(\ref{H_zeroset}) define
coisotropic deformations. One can easily check that the coisotropy
conditions for Hamiltonians~(\ref{H_zeroset}) exactly coincides
with~(\ref{omega_comp}). In particular, for polynomials
$\Omega_{n}(p_{1})$ of the type~(\ref{dKP_zeroset}) the
coisotropic deformations are given by the dKP hierarchy.

Reformulation of the dispersionless integrable equations and
hierarchies as the coisotropic deformations not only gives us a
technical simplification but also opens a way for essential
generalizations and a novel interpretation~\cite{KM}.

A way for considerable extension of the above approach lies in the
observation that one can pass from the zero set of the functions
$h_{n}$ to the ideal $J$ generated by them. Indeed, any polynomial
function $f_{n}$ of $h_{1}$, $h_{2},\dots$ belong to $\Gamma$ and
the coisotropy condition $\left.\left \{f_{n},f_{m}
\right\}\right|_{\Gamma} = 0$ is satisfied due to $\left.\left
\{h_{n},h_{m} \right\}\right|_{\Gamma} = 0$. Moreover, the both
sets of functions give rise to the same differential equations for
the potentials.

So the coisotropy condition takes the form of the closeness of $J$
with respect to the Poisson bracket~(\ref{poisson_defgen}), i.e.
\begin{equation}
\label{coisotropy cond} \left\{J,J \right\} \subset J.
\end{equation}
A passage to the ideal $J$ gives us the freedom to choose its basis
in different ways. Let us consider the ideal generater by the dKP
Hamiltonians $h_{0} \equiv -p_{0} + 1$ and
\begin{equation}
\label{H_ideal} h_{n} = - p_{n} + p^{n} + \sum_{m=0}^{n-2}
u_{nm}(x) p^{m}, \qquad n=1,2,3,\dots
\end{equation}
as an example, where for convenience we add Hamiltonians $h_{0} =
-p_{0}+1$ and $h_{1} = -p_{1} + p$. Formally, we have to introduce
the variable $x$ conjugate to $p$. This implies that $u_{nm}$ will
depend only on $x_{1}+x$. First, since $-p_{2} + p_{1}^{2} + u =0$
on $\Gamma$ one has $u = p_{2} - p_{1}^{2}$. Substituting this
expression for $u$ into $h_{3}$ , one gets a new Hamiltonian
\begin{equation}
\label{h3_new} \tilde{h}_{3} = - p_{3} - \frac{1}{2} p_{1}^{3} +
\frac{3}{2} p_{1} p_{2} + v.
\end{equation}
Continuing such procedure, one obtains an infinite set of
Hamiltonians $\tilde{h}_{n}$ which have a form of the sum of
polynomials in $p_{1}$, $p_{2},\dots$ and the functions
$u_{n}(x_{1},x_{2},\dots)$, namely,
\begin{equation}
\label{hn_new} \tilde{h}_{n} = n P_{n}(\tilde{p}) + u_{n}(x),
\qquad n=2,3,\dots
\end{equation}
where $\tilde{p} = \left(-p_{1},-\frac{1}{2}
p_{2},-\frac{1}{3}p_{3}, \dots \right)$, $u_{2} = u_{20} = u$,
$u_{3}= u_{30} = v$ and $P_{n}(t)$ are Schur's polynomials defined
by the formula $\exp \left(\sum_{n=1}^{\infty} z^{n} t_{n} \right)
=$$\sum_{m=0}^{\infty} z^{m} P_{m}(t)$. The Hamiltonians
$\tilde{h}_{n}$ together with $h_{0}$ and $h_{1}$ form a basis for
the same ideal $J$ generated by the functions
$h_{n}$~(\ref{H_ideal}).

For the Hamiltonians~(\ref{hn_new}) the coisotropy
condition~(\ref{coisotropy cond}) implies that
\begin{equation}
\label{coisotropy_conseq} \left\{\tilde{h}_{n},\tilde{h}_{m}
\right\} = \sum_{k=1}^{n-2} \frac{n}{k (n-k)} \der{u_{m}}{x_{k}}
\tilde{h}_{n-k} - \sum_{k=1}^{m-2} \frac{m}{k (m-k)}
\der{u_{n}}{x_{k}} \tilde{h}_{m-k}
\end{equation}
and the following equations are satisfied
\begin{align}
\label{coisotropy_conseqa} &\frac{n-1}{n} \der{u_{n}}{x_{m-1}}
= \frac{m-1}{m} \der{u_{m}}{x_{n-1}}, \\
\label{coisotropy_conseqb} &\der{u_{m}}{x_{n}} - \der{u_{n}}{x_{m}}
+ \sum_{k=1}^{m-2} \frac{m}{k (m-k)} u_{m-k} \der{u_{n}}{x_{k}} -
\sum_{k=1}^{n-2} \frac{n}{k (n-k)} u_{n-k} \der{u_{m}}{x_{k}} = 0
\end{align}
where $m,n=2,3,\dots$. One can check
that~(\ref{coisotropy_conseqa}) and~(\ref{coisotropy_conseqb}) are
equivalent to the standard dKP hierarchy.

An interesting basis arises if one shall try to convert the
basis~(\ref{H_ideal}) into that bilinear in $p_{1},$ $p_{2},$
$p_{3},\dots$ . Such a basis can be easily built in the following
way. From the first equation~(\ref{dKP_zeroset}) one has
$p_{1}^{2} = p_{2} - u$. Substituting this expression for
$p_{1}^{2}$ into the second equation~(\ref{dKP_zeroset}), one gets
the equation
\begin{equation}
\label{f12}- f_{12} \equiv p_{1} p_{2} - p_{3} + \frac{1}{2} u p_{1}
+ v = 0.
\end{equation}
Then, getting from the second equation~(\ref{dKP_zeroset}) the
expression $p_{1}^{3} = p_{3} - \frac{3}{2} u p_{1} - v$ and
substituting it and $p_{1}^{2} = p_{2} - u$ into
equation~(\ref{H_ideal}) with $n=4$, one arrives at the equation
\begin{equation}
\label{f13} -f_ {13} = p_{1} p_{3} - p_{4} + \left(u_{42} -
\frac{3}{2} u \right) p_{2} - v p_{1} + u_{40} - u u_{42} = 0.
\end{equation}
Repeating such a procedure, one gets the family of Hamiltonians of
the form~\cite{KM}
\begin{equation}
\label{fjk} -f_{jk} \equiv p_{j} p_{k} - \sum_{l=0}^{j+k}
C_{jk}^{l}(x) p_{l} = 0, \qquad j,k,=1,2,3,\dots
\end{equation}
where $C_{jk}^{l}$ are certain coefficients and we denote $p_{0}
\equiv 1$. The minus sign in the l.h.s. of the above formulas is
choosen in order to have the same definition of $f_{jk}$ as that in
\cite{KM}. The relation
\begin{equation*}
\label{fjk_struct} f_{jk} = h_{j} h_{k} - \sum_{l \geq 0}
C_{jk}^{l}(x) h_{l} + p_{j} h_{k} + p_{k} h_{j}
\end{equation*}
shows that the zero set of $h_{k}$ coincides with zero set of
$f_{jk}$. Direct but cumbersome calculations demonstrate that the
coisotropy conditions for the Hamiltonians $f_{jk}$ again give
rise to the dKP hierarchy.

So, the choice of the basis in the ideal $J$ does not affect the
coisotropy conditions. But, remarkably, the above freedom allows us
to reveal a deep algebraic roots of the dispersionless integrable
equations. Indeed, equations~(\ref{fjk}) look very much like the
realization of the table of multiplication for an associative
algebra in the basis formed by $p_{0},$ $p_{1},$ $p_{2},\dots$ .
This observation leads us to the possibility to treat the dKP
hierarchy as the coisotropic deformations of a commutative
associative algebra~\cite{KM}.

Namely, let we have a commutative associative algebra. We choose a
basis $p_{0} (\equiv 1),$ $p_{1},$ $p_{2}, \dots$ in the algebra
and so we have the corresponding table of multiplication
\begin{equation}
\label{table} p_{j} p_{k} = \sum_{l=0} C_{jk}^{l} p_{l}.
\end{equation}
The commutativity means that $C_{jk}^{l}=C_{kj}^{l}$ while the
associativity implies
\begin{equation}
\label{associativity_cond} \sum_{l=0} C_{jk}^{l} C_{lm}^{p} =
\sum_{l=0} C_{mk}^{l} C_{lj}^{p}.
\end{equation}
To consider deformations we assume that the structure constants
$C_{jk}^{k}$ depend on a set of variables $x_{1},$ $x_{2},$
$x_{3},\dots$ . To specify deformations we associate with the table
of multiplication~(\ref{table}) the set of quadratic Hamiltonians
\begin{equation}
\label{fjk_quad} f_{jk} = -p_{j} p_{k} + \sum_{l=0} C_{jk}^{l}(x)
p_{l},
\end{equation}
where $x_{j},p_{j}$ form pairs of canonically conjugate variables
in the symplectic manifold $M$ equipped with the Poisson
bracket~(\ref{poisson_defgen}).

Then we consider the polynomial ideal $J = \left< f_{jk} \right >$
generated by these Hamiltonians and the submanifold $\Gamma$
defined by
\begin{equation}
\label{submanifold} \Gamma = \left\{\left(x_{j},p_{j} \in M;
f_{jk} = 0 \right) \right\}.
\end{equation}
Finally, if the ideal $J$ is closed with respect to the Poisson
bracket, i.e.
\begin{equation}
\label{coisotropy condbis} \left \{J,J \right\} \subset J
\end{equation}
so that $\Gamma$ is a coisotropic submanifold of $M$, the
functions $C_{jk}^{l}(x)$ are said to define a coisotropic
deformation of the associative algebra defined
by~(\ref{table})~\cite{KM}.

One can show that the condition~(\ref{coisotropy condbis}) is
satisfied if the structure constants $C_{jk}^{l}$ obey the system
of equations
\begin{align}
\label{struct_cond} \sum_{s=1} \left(C_{sj}^{m}
\der{C_{lr}^{s}}{x_{k}} + C_{sk}^{m} \der{C_{lr}^{s}}{x_{j}} -
C_{sr}^{m} \der{C_{jk}^{s}}{x_{l}} - C_{sl}^{m}
\der{C_{jk}^{s}}{x_{r}} + \der{C_{jk}^{m}}{x_{s}} C_{jr}^{s} -
\der{C_{lr}^{m}}{x_{s}} C_{jk}^{s} \right) = 0
\end{align}
The system of equations~(\ref{associativity_cond})
and~(\ref{struct_cond}) completely defines coisotropic deformations
of the associative algebra~(\ref{table}) and it is a basic one in
our approach~\cite{KM}. For this reason we will refer to it as the
central system of the theory of coisotropic deformaitons.

We emphasize that the central system is the system of algebraic
and differential equations for the structure constants
$C_{jk}^{l}$ only. We shall see that this feature is an essential
advantage of the approach which we discuss here.

\section{dKP coisotropic deformations: tau function and Hirota's equations.}
\label{sec_dKP} A central problem concerning the central
system~(\ref{associativity_cond}),~(\ref{struct_cond}) is, of
course, the existence and meaning of its nontrivial solutions. A
solution is provided by the dKP hierarchy for which the structure
constants $C_{jk}^{l}$ can be reconstructed through the potentials
$u_{nm}(x)$ is~(\ref{H_ideal}) via the formula~(\ref{fjk}).

On the other hand an analysis of equations~(\ref{fjk})
shows~\cite{KM} that the dKP structure constants $C_{jk}^{l}$ have
the following general form
\begin{equation}
\label{dKP_struct_gen} C_{kj}^{l} = \delta_{k+j}^{l} + H_{j-l}^{k}
+ H_{k-l}^{j}
\end{equation}
where $k,j,l = 0,1,2,3,\dots$, $\delta_{k}^{l}$ is the Kroneker
symbol and $H_{j}^{k}$ are functions of $x_{1}, x_{2},\dots$ such
that $H_{0}^{k} = 0$ and $H_{j}^{k} = 0$ for $j \leq -1$. We will
show that the use of the general (unparametrized)
form~(\ref{dKP_struct_gen}) of the structure constants in the
central system will lead us to the existence of the tau-function and
a novel algebraic interpretation of the dispersionless Hirota's
equations~\cite{KM}.

The first step is to implement the associativity conditions on the
coefficients $H_{j}^{k}$. Direct substitution of
expressions~(\ref{dKP_struct_gen}) into equations~(\ref{H_ideal})
and the use of the identity
\begin{equation*}\label{identity}
\sum_{l=n-1}^{k-1}H_{k-l}^{i}H_{l-n}^{m}=
\sum_{l=n-1}^{k-1}H_{k-l}^{m}H_{l-n}^{i}  ,
\end{equation*}
shows that the structure constants $C_{jk}^{l}$ obey the
associative conditions if and only if the bracket
\begin{equation}
\label{brackets2} \left [ H,H \right ]_{ikm}:=
H_{m}^{i+k}-H_{m+k}^{i}-H_{i+m}^{k}
+\sum_{l=1}^{i-1}H_{i-l}^{k}H_{m}^{l}+\sum_{l=1}^{k-1}H_{k-l}^{i}H_{m}^{l}
-\sum_{l=1}^{m-1}H_{m-l}^{k}H_{l}^{i}
\end{equation}
vanishes identically for any choice of the indices
$(i,k,m)\in{N}$. An interesting consequence of this result can be
drawn by contraction. Indeed one may check that the above
equations imply the useful symmetry relations
\begin{equation}
\label{symmetry relations2} pH_{p}^{i} = iH_{i}^{p}  .
\end{equation}
Next, one has to implement the coisotropy conditions
\label{coisotropy conditions}. In terms of the coefficients
$H_{k}^{i}$ the ensuing equations are rather complicated. However a
closer scrutiny shows that they are simplified drastically on
account of the associativity conditions just obtained. Indeed one
can prove that the coisotropy conditions may be reduced to the
equations
\begin{equation}
\frac{\partial [ H,H ]_{l,n,i-j}}{\partial x_{k}} + \frac{\partial
[ H,H ]_{l,n,k-j}}{\partial x_{i}} - \frac{\partial [ H,H
]_{i,k,l-j}}{\partial x_{n}} - \frac{\partial [ H,H
]_{i,k,n-j}}{\partial x_{l}}  = 0 , \nonumber
\end{equation}
which are automatically fulfilled owing to the associativity
conditions, and to the \textit{linear} equations
\begin{equation}
\frac{\partial H_{p}^{i}}{\partial x_{l}} = \frac{\partial
H_{p}^{l}}{\partial x_{i}}  . \nonumber
\end{equation}

So, to summarize, the associativity and coisotropy conditions in
the case of the structure constants of the
form~(\ref{dKP_struct_gen}) are together equivalent to the set of
quadratic algebraic equations
\begin{equation}\label{associativity2}
[ H,H ]_{ikl}=0 ,
\end{equation}
and to the set of linear differential equations
\begin{equation}\label{exactness}
\frac{\partial H_{p}^{i}}{\partial x_{l}} = \frac{\partial
H_{p}^{l}}{\partial x_{i}}
\end{equation}
having the form of a system of conservation laws. The
equations~(\ref{associativity2}) and~(\ref{exactness}) give the
specific form of the \textit{central system} for the dKP hierarchy.
It encodes all the informations about the hierarchy. In particular
it entails that for any solution of the central system one has
\begin{equation}\label{brackets}
\{ f_{ik},f_{ln} \}=\sum_{s,t \geq 1  } K_{ikln}^{st} f_{st}
\end{equation}
where
\begin{equation*}
K_{ikln}^{st}= \left (\delta_{it}\frac {\partial }{\partial
x_{k}}+\delta_{kt}\frac{\partial }{\partial x_{i}} \right)
(H_{l-s}^{n}+H_{n-s}^{l}) - (\delta_{nt}\frac{\partial}{\partial
x_{l} }+\delta_{lt}\frac{\partial}{\partial x_{n}
})(H_{i-s}^{k}+H_{k-s}^{i})  .
\end{equation*}
From this formula one sees that the Hamiltonians $f_{jk}$ of the
dispersionless KP hierarchy form a Poisson algebra. The above
central system can be seen also as the dispersionless limit of the
central system of the full dispersive KP hierarchy~\cite{FM}.

There are presently two strategies to decode the informations
contained in the central system. According to the first strategy,
one first tackles the associativity
conditions~(\ref{associativity2}), noticing that they allow to
compute the coefficients $(H_{k}^{2},H_{k}^{3},\ldots)$ as
polynomial functions of $H_{k}^{1}$. For instance, the symmetry
conditions
\begin{equation*}
H_{1}^{2}=2H_{2}^{1}  \qquad   H_{1}^{3} = 3H_{3}^{1}
\end{equation*}
give $(H_{1}^{2},H_{1}^{3})$, and then the
condition~(\ref{associativity2}) with $i=k=1$, $l=2$, i.e.
\begin{equation*}
H_{1}^{3}-H_{3}^{1}-H_{2}^{2}+H_{1}^{1}H_{1}^{1} = 0
\end{equation*}
gives $H_{2}^{2}$, and so forth. Renaming the free coefficients as
suggested by the table of multiplication of the previous section,
that is by setting
\begin{equation*} H_{1}^{1}=-1/2u,  \qquad
H_{2}^{1}=-1/3v,  \qquad H_{3}^{1}=-1/4w+1/8u^{2} ,
\end{equation*}
one gets
\begin{equation*}
H_{1}^{2}=-2/3v,  \qquad  H_{2}^{2}=-1/2w+1/2u^{2}, \qquad
H_{1}^{3}=-3/4w+3/8u^2.
\end{equation*}
At this point one plugs these expressions into the simplest linear
coisotropy conditions
\begin{equation*}
\frac{\partial H_{1}^{1}}{\partial x_{2} }-\frac{\partial
H_{1}^{2}}{\partial x_{1} }=0, \qquad \frac{\partial
H_{2}^{1}}{\partial x_{2} }-\frac{\partial H_{2}^{2}}{\partial
x_{1} }=0, \qquad
\frac{\partial H_{1}^{1}}{\partial x_{3} }-\frac{\partial H_{1}^{3}}{\partial x_{1} }=0 ,\\
\end{equation*}
arriving to the equations
\begin{align*}
\frac{\partial v }{\partial x_{1} } &= 3/4\frac{\partial u}{\partial x_{2} }, \\
\frac{\partial v }{\partial x_{2} } &= 3/2\frac{\partial w
}{\partial x_{1} }-3u \frac{\partial u }{\partial x_{1}},
\\
\frac{\partial u }{\partial x_{3} } &=
3/2\frac{\partial w }{\partial x_{1} }-3/2u\frac{\partial u }{\partial x_{1} }  .\\
\end{align*}
The elimination of $\frac{\partial w }{\partial x_{1} }$ leads
finally to the dKP equation and to the higher equations, if one
insists enough in the computations. By this strategy one come back
to the hierarchy in its standard formulation.

A simple inversion of the order in which the equations are
considered leads instead to the Hirota's formulation. It is enough
to remark that equations (\ref{exactness}) entail the existence of a
sequence of potentials $S_{m}$ such that
\begin{equation*}
H_{m}^{i} = \frac{\partial S_{m} }{\partial x_{i} }  .
\end{equation*}
Then the symmetry conditions (\ref{symmetry relations2}), oblige
the potentials $S_{m}$ to obey the constraints
\begin{equation}
i\frac{\partial {S_{i}}}{\partial {x_{l}}}=l\frac{\partial
{S_{l}}}{\partial {x_{i}}}  , \nonumber
\end{equation}
which in turn entail the existence of a superpotential
$F(x_{1},x_{2},\ldots)$ such that
\begin{equation}
\label{superpotential} S_{i}=-1/i\frac{\partial {F}}{\partial
{x_{i}}},  \qquad H_{m}^{i}=-1/m\frac{\partial^{2}{F}}{\partial
{x_{i}}\partial {x_{m}}}.
\end{equation}
This result provides a second parametrization of the structure
constants, after that described before. The insertion of the new
parametrization into the full set of associativity conditions
\label{associativity} finally leads to the system of equations
\begin{gather}
\label{Hirota}
\begin{aligned}
&-\frac{1}{m}F_{i+k,m}+\frac{1}{m+k}F_{i,k+m}+\frac{1}{i+m}F_{k,i+m}\\
&+\sum_{l=1}^{i-1}\frac{1}{m(i-l)}F_{k,i-l}F_{l,m}+\sum_{l=1}^{k-1}\frac{1}{m(k-l)}F_{i,k-l}F_{l,m}\\
&-\sum_{l=1}^{m-1}\frac{1}{i(m-l)}F_{k,m-l}F_{i,l}=0  ,
\end{aligned}
\end{gather}
where $F_{i,k}$ stands for the second-order derivative of $F$ with
respect to $x_{i}$ and $x_{k}$. These equations are equivalent to
the celebrated Hirota's bilinear equations for the tau function of
the dispersionless KP hierarchy (see e.g.\cite{TT1,TT2, Mar1}). For
instance, for $(i=k=1,m=2)$ or
 $(i=m=1,k=2)$ or  $(m=k=1,i=2)$ one obtains directly the first Hirota's
equation
\begin{equation}
-1/2F_{2,2}+2/3F_{1,3}-(F_{1,1})^{2}=0  . \nonumber
\end{equation}
For $(i=1,k=2,m=2)$ or $(i=2,k=1,m=2)$ or $(i=1,k=1,m=3)$ it gives
instead the second Hirota's equation
\begin{equation}
1/2F_{1,4}-1/3F_{2,3}-F_{1,1}F_{1,2}=0  . \nonumber
\end{equation}
Higher equations~(\ref{Hirota}) do not separately coincide with
Hirota's bilinear equations, but are together equivalent to standard
Hirota's equations of the same weight.

Thus, the dispersionless Hirota's equations for the dKP hierarchy
is nothing but the associativity conditions~(\ref{associativity2})
under the parametrization~(\ref{superpotential}). We emphasize
that this result is due to the use of the structure
constants~(\ref{dKP_struct_gen}) in the central system instead of
potentials $u_{nm}$.

We note that the functions $K_{ikln}^{st}$ in the
formula~(\ref{brackets}) are linear combinations of third- order
derivatives ot the tau function $F$. For example,
\begin{equation}
\label{bracket_thirder} \{ f_{11},f_{1n} \} =
-2\sum_{s=1}^{s=n-1}\left( \frac{1}{(n-s)} \frac{\partial^3
F}{{\partial x_{1} }^{2} \partial x_{n-s} }\right)f_{1s} \quad
n=2,3,\ldots .
\end{equation}
It is, probably, not just a coincidence that the third order
derivatives of the tau function appear both in WDVV
equations~\cite{witten,Dij,Mar1} and as the structure constants in
equations~(\ref{bracket_thirder}).

We note also that the coisotropy
conditions~(\ref{bracket_thirder}) break at the points where the
third order derivatives of $F$ blow up. These points correspond to
the singular sector of the dKP hierachy.

\section{Extensions: dmKP, Harry Dym and d2DTL hierarchies.}
\label{sec_extensions} Several generalizations of the results
presented in the section~\ref{sec_dKP} are associated with rather
obvious extensions of the polynomial algebra defined by the
formula~(\ref{H_ideal}).

First extension is to relax the polynomials in the r.h.s
of~(\ref{H_ideal}) permitting them to contain the terms
$u_{n,n-1}(x) p^{n-1}$, i.e. to consider the family of
Hamiltonians
\begin{gather}
\label{h_extension}
\begin{aligned}
h_{0} &= -p_{0} + 1 \\
h_{1} &= -p_{1} + p \\
&\dots \\
h_{n} &= -p_{n} + \sum_{m=0}^{n} u_{nm}(x) p^{m},\qquad
n=2,3,4,\dots
\end{aligned}
\end{gather}
with $u_{nn} = 1$. The coisotropy condition again is given by the
equation $\left\{h_{n},h_{m} \right\} = 0$ where the Poisson bracket
is defined as
\begin{equation}
\label{poisson_defext} \left\{f,g \right\} = \der{f}{x} \der{g}{p}
- \der{f}{p} \der{g}{x} + \sum_{i=1} \left(\der{f}{x_{i}}
\der{g}{p_{i}} - \der{f}{p_{i}} \der{g}{x_{i}} \right).
\end{equation}
The variable $x_{0}$ conjugate to $p_{0}$ is a cyclic one and we
omit the corresponding term from~(\ref{poisson_defext}). In virtue
of the form of $h_{1}$ the coefficients $u_{nm}$ depend on the sum
$x+x_{1}$.

The coisotropy conditions for the Hamiltonians~(\ref{h_extension})
give rise to the generalised dKP hierarchy (see~\cite{KM}). In the
special gauge $u_{20} = 0,$ $u_{30}=0$ one gets the dmKP
hierarchy~\cite{KM}.

Further obvious extension is to allow the coefficients $u_{nn}$
depend on $x$. The corresponding deformations contain the
dispersionless Harry Dym hierarchy~\cite{Li} as a subclass.
Indeed, with the choice
\begin{gather}
\label{h_harrydym}
\begin{aligned}
h_{2} &= -p_{2} + \rho^{2} p_{1}^{2}, \\
h_{3} &= - p_{3} + \rho^{3} p_{1}^{3} + \mu \rho^{2} p_{1}^{2}
\end{aligned}
\end{gather}
one gets from the coisotropy condition the system
\begin{gather}
\label{coisotropy_harrydym}
\begin{aligned}
\der{\left(\rho^{2}\right)}{x_{3}} &= \der{\left(\mu \rho^{2}\right)}{x_{2}}, \\
\der{\mu}{x_{1}} &= \frac{3}{2} \der{\rho}{x_{2}}
\end{aligned}
\end{gather}
that is the dispersionless $2+1-$dimensional Harry Dym
equation~\cite{Li}.

In the above extensions one remains within the class of polynomial
algebras with the single generator $p$. A generalization of the
whole approach to polynomial algebras with $n$ generators via the
gluing process has been proposed in~\cite{KM}. In the simplest case
of two generators $p$ and $q$ it consists in adding to two
polynomial algebras~(\ref{h_extension}) the gluing relation
\begin{equation}
\label{pqgenerators} p q = a p + b q + c.
\end{equation}
The relation~(\ref{pqgenerators}) allows us to build the whole
table of multiplications between $p_{i}q_{k}$. It was shown
in~\cite{KM} that the corresponding coisotropic deformations are
given by the two-point dKP hierarchy which is equivalent to the
Whitham universal hierarchy on the Riemann sphere with two
punctures~\cite{Kri3}. In particular, it contains the simplest
$2+1-$dimensional Benney equation introduced in~\cite{Za3}.

Here we would like to indicate a way to incorporate the
dispersionless two-dimensional Toda lattice (d2DTL) in our scheme.
For this purpose we consider two general copies of polynomial
algebras and, hence, two families of Hamiltonians
\begin{gather}
\label{dTL_ham1}
\begin{aligned}
h_{0} &= -p_{0} + 1 \\
h_{1} &= -p_{1} + a p + b \\
&\dots \\
h_{n} &= -p_{n} + \sum_{m=0}^{n} u_{nm}(x,y) p^{m}
\end{aligned}
\end{gather}
and
\begin{gather}
\label{dTL_ham2}
\begin{aligned}
\tilde{h}_{0} &= h_{0} = -p_{0} + 1 \\
\tilde{h}_{1} &= -q_{1} + \tilde{a} q + \tilde{b} \\
&\dots \\
\tilde{h}_{n} &= -q_{n} + \sum_{m=0}^{n}
\tilde{u}_{nm}(x,y) q^{m}
\end{aligned}
\end{gather}
glued by the relation
\begin{equation}
\label{gluing} f = pq - 1 =0.
\end{equation}
Here $a$, $b$, $u_{nm}$, $\tilde{a}$, $\tilde{b}$, $\tilde{u}_{nm}$
are functions of the variables $x,$ $x_{2},$ $x_{2},\dots;$ $y,$
$y_{1,}$ $y_{2},\dots$ canonically conjugate to $p,$ $p_{1,}$
$p_{2},\dots;$ $q,$ $q_{1},$ $q_{2}, \dots$ .

Let us analyze now the coisotropy condition for these
Hamiltonians. It is not difficult to check that the use of the
canonical Poisson bracket of the type~(\ref{poisson_def}), for
instance, for the Hamiltonians $h_{1},$ $\tilde{h}_{1},$ $f$ gives
rise to a too strong constraints on $a,$ $b,$ $\tilde{a}, $
$\tilde{b}$ and consequently to trivial deformations.

If instead one uses a modified Poisson bracket defined as
\begin{gather}
\label{poisson_def_mod} \begin{aligned} \left\{f,g \right\} &= p
\left(\der{f}{x} \der{g}{p} - \der{f}{p} \der{g}{x} \right) - q
\left(\der{f}{y} \der{g}{q} - \der{f}{q} \der{g}{y} \right) \\
&+ \sum_{i=1} \left(\der{f}{x_{i}} \der{g}{p_{i}} - \der{f}{p_{i}}
\der{g}{x_{i}} +\der{f}{y_{i}} \der{g}{q_{i}} - \der{f}{q_{i}}
\der{g}{y_{i}} \right)
\end{aligned}
\end{gather}
the situation changes drastically. Indeed, from the conditions
$\left.\left\{h_{1},f \right\}\right|_{\Gamma} = 0$ and
$\left.\left\{\tilde{h}_{1},f \right\}\right|_{\Gamma} = 0$ one
obtains
\begin{gather}
\label{ab_eq}
\begin{aligned}
a_{x} - a_{y} &= 0, \qquad b_{x} - b_{y} = 0, \\
\tilde{a}_{x} - \tilde{a}_{y} &= 0, \qquad \tilde{b}_{x} -
\tilde{b}_{y} = 0,
\end{aligned}
\end{gather}
while the condition $\left.\left\{h_{1},\tilde{h}_{1}
\right\}\right|_{\Gamma} = 0$ gives
\begin{gather}
\label{ab_eq2}
\begin{aligned}
a_{y_{1}} + a \tilde{b}_{x} &= 0, \\
\tilde{a}_{x_{1}} - \tilde{a} b_{y} &= 0, \\
\tilde{a} a_{y} + a \tilde{a}_{x} + b_{y_{1}} - \tilde{b}_{x_{1}}
&= 0.
\end{aligned}
\end{gather}
At the particular gauge $b = -a$, $\tilde{b} = - \tilde{a}$ the
system~(\ref{ab_eq}),~(\ref{ab_eq2}) implies that
\begin{gather}
\label{ab_gauge}
\begin{aligned}
a_{y_{1}} - a \tilde{a}_{x} = 0, \\
\tilde{a}_{x_{1}} + \tilde{a} a_{x} = 0.
\end{aligned}
\end{gather}
From this system one gets
\begin{equation}
\label{dTL} \theta_{x_{1}y_{1}} + \left(e^{\theta}\right)_{xx} = 0
\end{equation}
where $\theta = \log \left(a \tilde{a} \right)$. It is the
Boyer-Finley or dispersionless two-dimensional Toda lattice (d2DTL)
equation (see e.g.~\cite{Kod2,TT2,Kri3}). Considering the coisotropy
conditions for the higher Hamiltonians $h_{n}$, $\tilde{h}_{n}$, one
will obtain the higher d2DTL equations and the whole d2DTL
hierarchy.

The fact that in order to get d2DTL as coisotropic deformations of
the polynomial algebras one has to consider the modified Poisson
bracket~(\ref{poisson_def_mod}) is of considerable importance. It
indicates that the symplectic structure of the deformations should
be consistent with the algebra to be deformed. Of course, one can
easily transform the modified bracket~(\ref{poisson_def_mod}) into a
canonical one introducing the variable $w$ such that $p=\exp (w)$,
$q = \exp (-w)$. But in terms of this variable the Hamiltonians
$h_{n}$, $\tilde{h}_{n}$ become the polynomials of exponentials
$\exp(w)$ and $\exp(-w)$. This phenomena and the first half of the
bracket~(\ref{poisson_def_mod}) (at $q = 1/p$) are well-known for
the d2DTL hierarchy (see e.g.~\cite{Kod2,TT2,Kri3}).

Similar situation of consistency of Poisson bracket and the form
of Hamiltonians take place for dispersionless equations considered
in~\cite{blas}.

\section{Coisotropic deformations of the Jordan's triple systems:
hierarchies of B type.} \label{sec_Btype} Quite different and
interesting structures arise for "algebras" compatible with a
discrete group, i.e. in the cases when discrete group acts
nontrivially  on a basis while the structure constants remain
invariant. Here we will consider only a simple example of the group
of simultaneous reflections of all elements of a basis $p_{i}
\rightarrow -p_{i}$.

Let us begin with the one component case. A natural basis now is
given by odd powers $p^{2i+1}$ and the analogs of the
polynomials~(\ref{H_ideal}) are of the form
\begin{gather}
\label{polyn_B}
\begin{aligned}
p_{1} &= p ,\\
p_{3} &= p^{3} + u p,\\
p_{5} &= p^{5} + v_{3} p^{3} + v_{1} p ,\\
p_{7} &= p^{7} + w_{5} p^{5} + w_{3} p^{3} + w_{1} p
\end{aligned}
\end{gather}
and so on. Such an algebra, obviously, is not closed under the
usual product operation $p_{i}p_{k}$. But the relations
\begin{gather}
\label{product_B}
\begin{aligned}
p_{1}^{3} & = p_{3} - u p_{1}, \\
p_{1}^{2} p_{3} & = p_{5} + \left(u - v_{3} \right) p_{3} -
\left[u \left(u-v_{3}\right) + v_{1} \right] p_{1}, \\
p_{1} p_{3}^{2} &= p_{7} + \left(2 u - w_{5} \right) p_{5} + \left
[u^{2} - w_{3} - v_{3} \left(2 u - w_{5} \right) \right] p_{3} \\
&+ \left[\left(2u -w_{5} \right) \left(v_{3} u - v_{1} \right) - u
\left( u^{2} - w_{3}\right) - w_{1} \right] p_{1}, \\
p_{1}^{2} p_{5} &= p_{7} + \left(v_{3} - w_{5} \right) p_{5} +
\left[v_{1} -w_{3} - v_{3} \left(v_{3} - w_{ 5} \right) \right]
p_{3}  \\
&+ \left[\left(v_{3} - w_{5} \right) \left(v_{3} u - v_{1} \right)
- u \left(v_{1} -w_{3} \right) - w_{1} \right] p_{1}
\end{aligned}
\end{gather}
and similar one suggest to consider the cubic operation
$p_{i}p_{k}p_{l}$. So, the odd polynomials~(\ref{polyn_B}) provide
us with an example of the ``algebra" with the basis $p_{1}$,
$p_{3}$, $p_{5}$, $\dots$ closed under the commutative trilinear
operation
\begin{equation}
\label{trilinear} p_{i} p_{k} p_{l} = \sum_{m=1}^{i+k+l}
C_{ikl}^{m} p_{m}
\end{equation}
where all indices take only odd integer values, and
\begin{align*}
C_{111}^{3} &= 1, \qquad &C_{111}^{1} = - u, \qquad C_{113}^{5} =
1, \\
C_{113}^{3} &= u - v_{3}, \qquad &C_{113}^{1} = - \left[u \left(u
- v_{3} \right) + v_{1} \right]
\end{align*}
and so on. Algebraic structures defined by the trilinear
law~(\ref{trilinear}) are known as the Jordan's triple systems
(see e.g.~\cite{jac,neh}). The associativity for the Jordan's
triple system is defined as
\begin{equation}
\label{triple_associativity} \left(p_{i} p_{k} p_{l} \right) p_{n}
p_{m} = p_{i} p_{k} \left(p_{l} p_{n} p_{m} \right) = p_{i}
\left(p_{k}p_{l} p_{n} \right) p_{m}
\end{equation}
that leads to the following ``associativity" conditions
\begin{gather}
\label{triple_associativity_cond}
\begin{aligned}
\sum_{t} C_{ikl}^{t} C_{tnm}^{s} = \sum_{t} C_{lnm}^{t}
C_{ikt}^{s}, \\
\sum_{t} C_{lnm}^{t} C_{ikt}^{s} = \sum_{t} C_{kln}^{t}
C_{imt}^{t}.
\end{aligned}
\end{gather}
One can apply the approach discussed above to define coisotropic
deformations of Jordan's triple systems. So, we convert the
multiplication table~(\ref{trilinear}) into to the zero set $\Gamma$
of the Hamiltonians
\begin{equation}
\label{fikl_triple} f_{ikl} := - p_{i} p_{k} p_{l} + \sum_{m=1}
C_{ikl}^{m}(x) p_{m}, \qquad i,k,l = 1,3,5,\dots
\end{equation}
and then demand the coisotropy of $\Gamma$ with respect to the
canonical Poisson bracket. One gets the coisotropy deformations
equations for the structure constants $C_{ikl}^{m}$ which are
similar to the equations~(\ref{struct_cond}).

Here we will consider the simplest of them which arise as the
coisotropy conditions for the lowest Hamiltonians
\begin{equation*}
\label{f111_triple} f_{111} = - p_{1}^{3} + p_{3} - u p_{1} \qquad
\textup{and} \qquad f_{113} = -p_{1}^{2} p_{3} + p_{5} - v p_{3} - w
p_{1}.
\end{equation*}
One gets
\begin{gather}
\label{f111_f113_bracket}
\begin{aligned}
\left\{f_{111},f_{113} \right\} &= \left(-u_{x_{2}} + 3 w_{x_{1}}
\right) f_{111} + \left(- 2 u_{x_{1}} + 3 v_{x_{1}} \right)
f_{113} + \left(-2 u_{x_{1}} + 3 v_{x_{1}} \right) p_{5} \\
&+ \left(-u_{x_{3}} + 3 w_{x_{1}} + 2 v u_{x_{1}} - 3 v v_{x_{1}}
- v_{x_{3}} + u v_{x_{1}} \right) p_{3} \\
&+ \left(u u_{x_{3}} - 2 u w_{x_{1}} + 2 w u_{x_{1}} - 3 w
v_{x_{1}} - w u_{x_{1}} - u_{x_{2}} v + u_{x_{5}} - w_{x_{3}}
\right) p_{1}.
\end{aligned}
\end{gather}
The r.h.s. of~(\ref{f111_f113_bracket}) vanishes on $\Gamma$ if
\begin{align*}
2 u_{x_{1}} - 3 v_{x_{1}} &= 0,\\
u_{x_{3}} - 3 w_{x_{1}} + v_{x_{3}} - u v_{x_{1}} &= 0, \\
u_{x_{5}} + u u_{x_{3}} - 2 u w_{x_{1}} - u_{x_{1}} w - v
u_{x_{3}} - w_{x_{3}} &=0.
\end{align*}
From the first equation as usual one has $v = (2/3) u$. So, one gets
the following system
\begin{align*}
w_{x_{1}} + \frac{1}{9} \left(u \right)_{x_{1}}^{2} - \frac{5}{9}
u_{x_{3}} &= 0, \\
u_{x_{5}} - \frac{7}{9} u u_{x_{3}} + \frac{4}{9} u^{2} u_{x_{1}}
- u_{x_{1}} w - w_{x_{3}} &=0
\end{align*}
or, equivalently, the equation
\begin{equation}
\label{dKP_B} \frac{9}{5} u_{x_{5}} - u u_{x_{3}} + u^{2}
u_{x_{1}} - u_{x_{1}} \partial^{-1}_{x_{1}} u_{x_{3}} -
\partial^{-1}_{x_{1}}u_{x_{3}x_{3}} = 0.
\end{equation}
It is just the dKP equation of the B type~\cite{tak,K2}. In a
similar manner one can gets the whole dBKP hierachy which represent
itself the coisotropic deformations of the Jordan's triple
systems~(\ref{trilinear}).

The form of the structure constants analogous
to~(\ref{dKP_struct_gen}) of dKP case is the following
\begin{gather}
\label{dKP_ struct_gen_B} \begin{aligned}C_{ikl}^{m} &=
\delta_{i+k+l+1}^{m} + H _{2(k+l-m)+1}^{2i+1} +
H_{2(i+l-m)+1}^{2k+1} + H_{2(i+k-m)+1}^{2l+1}
\\&+
 \simbon{\sum_{p,t =0}^{\infty}}{p+t=l-1-m; \: m\geq 0 }H_{2p+1}^{2i+1}
 H_{2t+1}^{2k+1} \;\;+ \simbon{\sum_{p,s =0}^{\infty}}{p+s=k-1-m; \: m\geq 0 }H_{2p+1}^{2i+1}
 H_{2s+1}^{2l+1} \\&+ \;\; \simbon{\sum_{t,s =0}^{\infty}}{t+s=i-1-m; \: m\geq 0 }H_{2t+1}^{2k+1}
 H_{2s+1}^{2l+1}
\end{aligned}
\end{gather} where $H_{2p+1}^{2i+1} = 0$ at $p\leq 1$. The associativity
conditions~(\ref{triple_associativity_cond}) are equivalent to
infinite set of cubic equations for $H_{2p+1}^{2i+1}$ and the
coisotropy conditions are equivalent to these algebraic
associativity equations and the exactness equations similar to the
dKP case.

The two point dKP hierarchy with the involution $p_{i} \rightarrow
-q_{i}$, $q_{i} \rightarrow -q_{i}$ provides us with one more
interesting example. In this case one has two families of
relations~(\ref{polyn_B}), i.e.
\begin{gather}
\label{polyn_B_dKP}
\begin{aligned}
p_{2i+1} &= \sum_{k=0}^{i} v_{i,k} (x,y) p_{1}^{2k+1}, \\
q_{2i+1} &= \sum_{k=0}^{i} w_{i,k} (x,y) q_{1}^{2k+1}
\end{aligned}
\end{gather} glued by
\begin{equation}
\label{gluing_triple} p_{1} q_{1} = u(x,y)
\end{equation}
where $(x,y) = \left(x_{1},x_{3},x_{5},\dots;
y_{1},y_{3},y_{5},\dots \right)$.

The corresponding closed ``algebra" is defined by the following
table of multiplication
\begin{gather}
\label{table_triple}
\begin{aligned}
p_{i}p_{k}p_{l} &= \sum_{m=1} A_{ikl}^{m}(x,y) p_{m}, \\
q_{i}q_{k}q_{l} &= \sum_{m=1} B_{ikl}^{m}(x,y) q_{m}, \\
p_{i}p_{k}q_{l} &= \sum_{m=1} C_{ikl}^{m}(x,y) p_{m} + \sum_{m=1}
D_{ikl}^{m}(x,y) q_{m},  \\
p_{i}q_{k}q_{l} &= \sum_{m=1} E_{ikl}^{m}(x,y) p_{m} + \sum_{m=1}
F_{ikl}^{m}(x,y) q_{m}.
\end{aligned}
\end{gather}
The formulas~(\ref{table_triple}) define the Jordan's double
triple system. In the
parametrization~(\ref{polyn_B_dKP}),~(\ref{gluing_triple}) the
structure constants $A$ and $B$ are given by the relations of the
type~(\ref{product_B}) while
\begin{gather}
\label{struct_triple}
\begin{aligned}
C_{111}^{1} =
 &\:u, \qquad D_{111}^{m} = 0,\qquad E_{111}^{m} =0,
\qquad F_{111}^{1} = u, \\
C_{113}^{1} = w_{1,1}& u, \qquad D_{113}^{1} = u^{2}, \qquad
E_{311}^{1} = u^{2}, \qquad F_{311}^{1} = u v_{1,1},
\end{aligned}
\end{gather}
and so on.

In general, the ``structure constants" $A$, $B$, $C$, $D$, $E$,
$F$ obey the set of cubic ``associativity" conditions. The
coisotropy condition for the Hamiltonians defined
by~(\ref{table_triple}) again give rise to the integrable
deformations of the triple system~(\ref{table_triple}). It can be
called the two-point dBKP hierarchy.

Let us consider the set of three lowest Hamiltonians
from~(\ref{polyn_B_dKP}),~(\ref{gluing_triple}), i.e.
\begin{gather}
\label{H_triple}
\begin{aligned}
f &= p_{3} - p_{1}^{3} + v p, \\
\tilde{f} &= q_{3} - q_{1}^{3} + w q_{1}, \\
H & = p_{1} q_{1} - u(x,y).
\end{aligned}
\end{gather}
The coisotropy conditions for the zero set $\Gamma : f = g = H =
0$ take the form
\begin{gather}
\label{coisotropy cond_triple} \begin{aligned} \left \{f,
\tilde{f}  \right\} &= 3  \left (w_{x_{1}} p_{1} -
v_{y_{1}} q_{1} \right ) H, \\
\left \{f, H  \right\} &= v_{x_{1}} H, \\
\left \{\tilde{f}, H  \right\} & = w_{y_{1}} H
\end{aligned}
\end{gather}
together with the equations
\begin{gather}
\label{coisotropy cond_aux}
\begin{aligned}
u_{x_{3}} + \left(u v \right)_{x_{1}} &= 0, \\
v_{y_{1}} - 3 u_{x_{1}} &= 0, \\
u_{y_{3}} + \left(u w \right)_{y_{1}} &= 0, \\
w_{x_{1}} - 3 u_{y_{1}} & = 0.
\end{aligned}
\end{gather}
This system represent the dispersionless limit of the
Nizhnik-Veselov-Novikov (dNVN) equation [16] (more precisely for
solutions of the form $u =$ $u (x_{1},y_{1},x_{3}+y_{3})$). The
whole family of the coisotropic deformations for the
algebra~(\ref{table_triple}) is given by the dNVN hierarchy.

Central systems for the algebraic structures discussed above,
existence of tau-functions and corresponding dispersionless Hirota's
equations will be considered elsewhere.

\section{Stationary reductions of dispersionless hierarchies and
 dynamical systems integrable on
 a fixed energy level.}
\label{sec_stationary} Integrable hierarchies and the whole
construction considered in this paper admit reductions for which
some variables $x_{i}$ are cyclic one. For example, for the dKP case
the cyclicity of the variable $x_{2}$ reduces it to the dKdV
hierarchy. Under such a reduction $\left\{p_{i}, f \right\} = 0$ for
any function $f$ and hence $p_{i} = $const. So the number of
canonical variables is reduced effectively by two and the number of
Hamiltonians becomes equal to the one-half of total number of
variables.

Let us analyze such a situation more carefully. We begin with the
dKP equation and assume that the variable $x_{3}$ is cyclic.
Hence, $p_{3} = $const. For this stationary reduction the dKP
equation becomes
\begin{equation}
\label{dKP_stat} \dersec{u}{x_{2}} + \dersec{\left(u^{2}
\right)}{x_{1}} = 0
\end{equation}
and one has two Hamiltonians~(\ref{dKP_zeroset})
\begin{gather}
\label{dKP_zeroset_stat}
\begin{aligned}
H &:= h_{2} = -p_{2} + p_{1}^{2} + u(x_{1},x_{2}), \\
H_{1} &:= h_{3} = p_{1}^{3} + \frac{3}{2} u p_{1} + v +
\textup{const}
\end{aligned}
\end{gather}
where $\der{v}{x_{1}} = \frac{4}{3} \der{u}{x_{2}}$. These
Hamiltonians are in involution $\left \{H,H_{1} \right\} = 0$.

For the Hamiltonian $H$ the associated dynamical system contains,
in particular, the equations
\begin{gather}
\label{dKP_stat_dynamical}
\begin{aligned}
\der{x_{1}}{t} &= 2 \der{H}{p_{1}} = 2 p_{1}, \\
\der{p_{1}}{t} & = - \der{H}{x_{1}} = - \der{u}{x_{1}},
\end{aligned}
\end{gather}
and, hence,
\begin{equation}
\label{dKP_stat_dynamical2} \dersec{x_{1}}{t} = -2 \der{u}{x_{1}}.
\end{equation}
For the dynamical system~(\ref{dKP_stat_dynamical}) the function
$H_{1}$ is the integral of motion cubic in momentum. This
observation reproduces the result of the paper~\cite{koz} that the
dynamical system~(\ref{dKP_stat_dynamical})
and~(\ref{dKP_stat_dynamical2}) with the potential
$u\left(x_{1},x_{2} \right)$ which obey equation~(\ref{dKP_stat})
has the additional integral of motion cubic in $p_{1}$.

In
 the case of cyclicity with respect to the variable $x_{n}$
($n\geq 4$) one has the Hamiltonian $H_{n} = p_{1}^{n} + \dots$, in
involution with $H$, which is the integral of motion for the
system~(\ref{dKP_stat_dynamical}) of the order $n$ (see
again~\cite{koz}).

Another simple example is provided by the stationary dNVN
system~(\ref{H_triple}), (\ref{coisotropy cond_triple}). In order
to establish the connection with the standard two-dimensional
dynamical systems we impose the constraints $y_{1} = \bar{x}_{1}$,
$q_{1} = \bar{p}_{1}$, $q_{3} = \bar{p}_{3}$, $w = \bar{v}$ where
bar means a complex conjugation, and assume that $u$ dependes only
on $x_{1}$, $\bar{x}_{1}$ and $t = x_{3} + y_{3}$. Passing to new
variables $x_{1} \rightarrow z := x_{1} + i x_{2}$, $p_{1}
\rightarrow p:= p_{1} + i p_{2}$, one rewrites the Hamiltonian $H$
in the form
\begin{equation}
\label{H_dVN} H = p_{1}^{2} + p_{2}^{2} + u \left(x_{1},x_{2},t
\right)
\end{equation}
while the corresponding equation looks like
\begin{gather}
\label{dVN}
\begin{aligned}
&\der{u}{t} + \der{\left(u v \right)}{z} + \der{\left(u \bar{v}
\right)}{\bar{z}}=0, \\
&\der{v}{\bar{z}} = 3 \der{u}{z}.
\end{aligned}
\end{gather}
It is the dispersionless Veselov-Novikov (dVN)
equation~\cite{Kri1,K3}. Instead of two Hamiltonians $f$ and
$\tilde{f}$~(\ref{H_triple}) one has for equation~(\ref{dVN})
their sum
\begin{equation}
\label{H_sum} \widetilde{H} = f+\tilde{f} = p_{3} + \bar{p}_{3} -
p^{3} + v p - \bar{p}^{3} + \bar{v} \bar{p}
\end{equation}
and
\begin{equation}
\label{bracker_sum} \left\{H,\widetilde{H} \right\} = \left(v_{z}
+ \bar{v}_{\bar{z}}  \right) H.
\end{equation}
The formulae~(\ref{H_sum}-\ref{bracker_sum}) mean that for any
solution of the dVN equation~(\ref{dVN}) the dynamical system with
the Hamiltonian $H$~(\ref{H_dVN}) possesses on the level of zero
energy $E = 0$ the additional integral of motion
$\widetilde{H}$~(\ref{H_sum}). In particular, in the stationary
case $\der{u}{t} = 0$, then $p_{3} + \bar{p}_{3}$ is a constant
and hence
\begin{equation} \label{add_int} \widetilde{H} = - p^{3} + v p -
\bar{p}^{3} + \bar{v} \bar{p} + \textup{const}.
\end{equation}
So, in this case one has the dynamical system with two degrees of
freedom which has at the zero energy level $E=0$ the additional
cubic integral of motion~(\ref{add_int}) if the potential $u$ is a
solution of the stationary $dVN$ equation
\begin{equation}
\label{dVN_stat} \der{\left(u v \right)}{z} + \der{\left(u \bar{v}
\right)}{\bar{z}} = 0, \qquad \der{v}{\bar{z}} = 3 \der{u}{z}.
\end{equation}
Such integrals of motion are usually called conditional or
configurational integrals of motion (see e.g.~\cite{bir,whit}).
Within a different approach the connection between the existence of
the conditional cubic integral~(\ref{add_int}) and
equation~(\ref{dVN_stat}) has been established in~\cite{hiet}
(formula (7.5.12)). Similarly the higher odd order conditional
integrals for the Hamiltonian system~(\ref{H_dVN}) found
in~\cite{hiet} are the Hamiltonians $H_{n}$ associated with higher
stationary dVN equations.

So, stationary dKP, stationary dVN hierarchies and other stationary
dispersionless integrable hierarchies provide us with the dynamical
systems completely Liouville integrable on the fixed energy level.

Full dispersionless hierarchies in our scheme are characterized by
the existence of number of independent Hamiltonians which is equal
to the one-half of dimension of the symplectic space minus one. They
represent the cases closest to the Liouville completely integrable
ones and one may call them next to the Liouville completely
integrable.
\\
\\
{\bf Acknowledgements.} The authors are grateful to A. Moro for
the help in preparation of the paper. The work has been partly
supported by the grants COFIN 2004 "Sintesi", and COFIN 2004
"Nonlinear Waves and Integrable systems"

\end{document}